\documentclass[twocolumn,showpacs,prb]{revtex4}

\usepackage{epsfig,wrapfig}
\usepackage{graphicx}
\usepackage{bm}

\newcommand{\beq}{\begin{equation}}
\newcommand{\eeq}{\end{equation}}

\begin{document}

\title{Non-Fermi-Liquid Behavior from the Fermi-Liquid Approach}
\author{
V.~A.~Khodel,$^{1,2}$ J.~W.~Clark,$^1$ H.~Li,$^1$ V.~M.~Yakovenko,$^3$
 and M.~V.~Zverev$^2$  } \vskip 0.8 cm
\affiliation {%
$^1$McDonnell Center
for the Space Science and Department of Physics, \\
Washington University, St.Louis, MO 63130, USA \\
$^2$Russian
Research Center Kurchatov Institute,
    Moscow, 123182 Russia \\
$^3$Condensed Matter Theory Center
and Center for Superconductivity Research, \\
Department of Physics, University of Maryland, \\
College Park, Maryland 20742-4111, USA
 } \vskip 0.3 cm

\vskip 0.5cm
\date{\today}

\begin{abstract}
Non-Fermi liquid behavior of strongly correlated Fermi systems is
derived within the Landau approach.  We attribute this behavior to
a phase transition associated with a rearrangement of the Landau
state that leads to flattening of a portion of the single-particle
spectrum $\epsilon({\bf p})$ in the vicinity of the Fermi surface.
We demonstrate that the quasiparticle subsystem responsible for
the flat spectrum possesses the same thermodynamic properties as a
gas of localized spins.  Theoretical results compare favorably
with available experimental data. While departing radically from
prevalent views on the origin of non-Fermi-liquid behavior, the
theory advanced here is nevertheless a conservative one of in
continuing to operate within the general framework of Landau
theory.

\end{abstract}


\maketitle

\section{Introduction}

This year is a jubilee for Fermi liquid (FL) theory. Fifty years
ago L.~D.~Landau published an article \cite{lan} devoted to
evaluation of thermodynamic properties of three-dimensional (3D)
liquid $^3$He, in which he postulated a one-to-one correspondence
between the totality of real, decaying single-particle (sp)
excitations of a Fermi liquid and a system of immortal interacting
quasiparticles. Two distinguishing features specify this system:
(i) its quasiparticle number is equal to the particle number (the
so-called  Landau-Luttinger theorem), and (ii) its entropy $S$,
given by the ideal-Fermi-gas combinatorial expression, coincides
with the entropy of the actual helium system. Treating the
ground-state energy $E_0$ as a functional of the quasiparticle
momentum distribution $n(p)$, Landau derived a formula
\begin{equation}
n(p,T)=\left[ 1+e^{\epsilon(p)/T}\right]^{-1}
\label{dist}
\end{equation}
for the quasiparticle momentum distribution $n(p,T)$ that resembles
the corresponding Fermi-Dirac formula for the ideal Fermi gas. However,
the quasiparticle energy $\epsilon(p)=\delta \Omega/\delta n(p)$ is
a variational derivative of the thermodynamic potential
$\Omega=E-\mu N$, where $E$ is the ground-state energy and $\mu$,
the chemical potential, with
respect to the momentum distribution $n(p)$, rather than the bare
sp energy $\epsilon^0_p$ entering  the respective Fermi-Dirac
formula. As we shall see, in strongly correlated Fermi systems,
this noncoincidence is crucial.

Landau's quasiparticle pattern of low-temperature phenomena in
Fermi liquids is universally recognized as a cornerstone of condensed-matter
theory.  However, a qwirk of fate is that although originally FL theory
was created for the explanation of  properties of 3D liquid $^3$He,
discrepancies between theoretical predictions and experimental data on
the low-T spin susceptibility $\chi(T)$ and the ratio $C(T)/T$ of the
specific heat $C(T)$ to temperature $T$ (both these must be constant
in FL theory)
first came to right in this liquid. The deviations, rather small at
extremely low $T$, increase with temperature.
In their book, Nozi\`eres and Pines\cite{pines} attribute
the departures to the damping of sp excitations.  During more than
thirty years this claim was regarded as a terminal diagnosis of FL theory.
However, the situation was turned upside down about ten years ago when
2D liquid $^3$He was studied quantitatively in a region of densities
$\rho$ where this system gradually evolves from a weakly correlated
gas to a strongly correlated liquid.\cite{godfrin,saunders}
The experimental data turned out to be even more challenging to
interpret than those for the 3D counterpart. Indeed, in the
density region where strong correlations set in, departures from
predictions of FL theory exhibit themselves more strongly,
the {\it lower} the temperature. This definitely rules out
damping of sp excitations as the cause for failure of FL theory,
since damping vanishes as $T\to 0$.

The departures in question occur in the density region where
the FL effective mass $M^*(\rho)$, extracted from the 2D
liquid $^3$He specific heat data \cite{saunders} using the
FL formula $C_{\mbox{\scriptsize FL}}(T)/T=p_FM^*/3$, is found
to be enhanced, with the enhancement rapidly increasing as $\rho$
approaches a critical value $\rho_s\simeq 0.072$ \AA$^{-2}$,
beyond which 2D liquid $^3$He solidifies.  Significantly,
the behavior of $\chi(T)$ changes drastically well before
$\rho$ reaches $\rho_s$: the product $\chi(T) T$
ceases to vanish at $T\to 0$, in agreement with the Curie law and
in contrast to predictions of standard FL theory. In doing so, the value
of this product {\it gradually increases} as $\rho\to \rho_s$.
Analogous behavior has been observed for other strongly correlated
Fermi liquids (see e.g. Ref.~\onlinecite{reznikov}).

Various theories have been suggested in attempts to identify
the source of non-Fermi-liquid (NFL) behavior in strongly correlated
Fermi systems.  Many of them are based on a theory advanced in
Refs.~\onlinecite{hertz,millis}, in which the NFL behavior is attributed
to spin fluctuations.  However, the spin-fluctuation scenario
fails to explain experimental results for thermodynamic properties
of heavy-fermion metals in the vicinity of the so-called quantum
critical point, especially in the presence of external
magnetic fields.\cite{coleman1,steglich,gegenwart,takahashi}

The gradually emerging Curie behavior of $\chi(T)$ in 2D liquid
$^3$He is sometimes attributed to a localization phase transition.
Unfortunately for this explanation, a homogeneous system without
impurities has no more chance to be slightly localized than
a women has to be slightly pregnant.  The Curie-Weiss behavior
of $\chi(T)$ in heavy-fermion metals is ordinarily described within
the Anderson $s{-}d$ model, in which localized $d$ states acquire
a nonzero spin as a result of interaction with delocalized $s$
electrons.  However, within this scenario it is hard to reconcile
the localization of electrons with Cooper pairing, which is
found to be unexpectedly strong in some cases.\cite{thompson}

We shall argue that the occurrence of the Curie-like term in
low-temperature behavior of the magnetic susceptibility $\chi(T)$
of a homogeneous Fermi liquid, which {\it gradually evolves under change
of input parameters}, is a signature of a phase transition known by
fermion condensation.  This phase transition is associated with
a rearrangement of {\it single-particle degrees of freedom}, rather
than collective ones.

\section{Rearrangement of sp degrees of freedom in finite Fermi systems}

It is instructive to begin studying such a rearrangement in
finite Fermi systems, in which damping of single-particle
excitations does not occur.  This closes off one favorite
escape route of critical readers and referees.  It is the
conventional wisdom of textbooks that under variation of input
parameters, two sp levels may repel or cross one other.  However,
as we shall demonstrate, the familiar dichotomy misses a further
alternative: levels can in fact merge.\cite{haochen} This
phenomenon is made possible by the variation of sp energies
with level occupation numbers -- a property central to Landau theory.
A primary condition for merging to occur is that the Landau-Migdal
interaction function $f$ is repulsive in coordinate space,\cite{haochen}
which holds for the effective $nn$ and $pp$ interactions in
the nuclear interior \cite{migdal} and for the electron-electron
interaction in atoms.

Consider a schematic model involving three equidistant neutron levels,
separated by an energy distance $D$ in an open shell of a spherical
nucleus. The levels are denoted $-$, $0$, and $+$, in order of
increasing energy. The sp energies $\epsilon_{\lambda}$ and wave
functions $\varphi_{\lambda}({\bf r})=R_{nl}(r)\Phi_{jlm}({\bf n})$
are solutions of
\begin{equation}
[p^2/2M+\Sigma({\bf r},{\bf p})]\varphi_{\lambda}({\bf r}) =
\epsilon_{\lambda}\varphi_{\lambda}({\bf r})\,,
\end{equation}
where $\Sigma$ stands for the self-energy.  In even-even spherical
nuclei, which in their ground states have total angular momentum
$J=0$ due to pairing correlations, the energies $\epsilon_{\lambda}$
are independent of the magnetic quantum number $m$ associated
with the total sp angular momentum $j$. We suppose that the
level $-$ is filled, the level $+$ is empty, and $N$ neutrons
are added to the level $0$, changing the density $\rho(r)$ by
$\delta\rho(r)=NR^2_{n_0l_0}(r)/4\pi$.

In what follows, we shall retain only a major, spin- and
momentum-independent part $V$ of the self-energy $\Sigma$ and
a primary, $\delta(r)$-like portion of the Landau-Migdal
interaction function $f$.  Accordingly, the FL relation between
$\Sigma $ and $\rho$ responsible for the variation
of $\epsilon_{\lambda}(n)$ with $n$ reduces to\cite{migdal}
\begin{equation}
  \delta V(r)=f[\rho(r)]\delta \rho(r) \ .
\label{rel1}
\end{equation}

When particles are added to the system, all energy levels are
shifted somewhat, but the level that receives the particles is
affected more strongly than the others. For the sake of
simplicity, the diagonal and nondiagonal matrix elements of $f$
are assigned the respective values
$$u=\int
R_{nl}^2(r)f\left[\rho(r)\right]R^2_{nl}(r)r^2dr/4\pi \ , $$
\vspace{-.5truecm}
\begin{equation} w= \int
R_{nl}^2(r)f\left[\rho(r)\right] R^2_{n_1l_1}(r)r^2dr/ 4\pi \,,
\label{mel}
\end{equation}
independently of the quantum numbers $nl,\,n_1l_1$.

Based on these assumptions and results, the dimensionless shifts
$\xi_k(N)=\left[\epsilon_k(N)-\epsilon_k(0)\right]/D$
for $k=0,+,-$ are given by
\begin{equation}
  \xi_0(N)=n_0U\ , \quad \xi_+(N)=\xi_-(N)=n_0W \  ,
\label{en1}
\end{equation}
where $n_k=N_k/(2j_k+1)$ is the occupation number of the level $k$,
$U=u(2j_0+1)/D$, and $W=w(2j_0+1)/D$. It is readily verified that
if $fp_FM/\pi^2\sim 1$, where $p_F=\sqrt{2M\epsilon_F}$ and $\epsilon_F$
is the Fermi energy, then the first of the integrals (\ref{mel})
has a value $u\simeq \epsilon_F/A$ and therefore $U\sim 1$, since
$D\sim \epsilon_F/A^{2/3}$ in spherical nuclei.

According to Eqs.~(\ref{en1}) at $(U-W)>1$, the difference
$d(N)=1+\xi_+(N)-\xi_0(N)$ changes sign at $n_{0c}=1/(U-W)$, before
filling of the level $+$ is complete.
At $n_0>n_{0c}$, in the standard scenario provided by Hartree-Fock
theory, all added quasiparticles must resettle into the  empty
sp level $+$. However, not all of the quasiparticles can take part
in the migration
process, since the situation would then be reversed, and the
roles of the levels interchanged: the formerly empty level,
lying above the formerly occupied one, would have the maximum
positive energy shift, rendering migration impossible.
Thus, the standard Fermi-liquid filling scenario, which prescribes
that {\it one and only one} sp level lying exactly at the Fermi
surface can remain unfilled, while all others must be completely
occupied or empty, encounters a catastrophe.

This catastrophe is resolved as follows \cite{ks,haochen}. Migration
occurs until the sp energies of the two levels in play coincide.
As a result, both of the levels, $0$ and $+$, become {\it partially}
occupied -- an impossible situation for the standard Landau state.
Solution of the problem reduces
to finding the minimum of the relevant energy functional
\begin{equation}
E_0=\epsilon_0(0)N_0+\epsilon_+(0)N_++\frac{1}{2}\left[u(N^2_0+N^2_+)
+2wN_0N_+\right]
\label{energy}
\end{equation}
with $N_k=\sum_m n_{km}$, through a variational condition
\begin{equation}
\frac{\delta E_0}{\delta n_{0m}}=\frac{\delta E}{\delta n_{+m_1}}=\mu \,,
\qquad
\forall m, m_1 \,,
\label{var}
\end{equation}
where $\mu$ is the chemical potential.  Such a condition  first appeared
in Ref.~\onlinecite{ks}, where homogeneous Fermi systems were addressed
without attention to the degeneracy of sp levels.
Eqs.~(\ref{var}) are conveniently rewritten as conditions
$$
\epsilon_0(N)=\epsilon_0(0)+N_0u+N_+w=\mu \ , $$
\vspace{-.5truecm}
\begin{equation}
\epsilon_+(N)=\epsilon_+(0)+N_0w+N_+u=\mu \label{en3}
\end{equation}
for coincidence of the sp energies $\epsilon_0$ and $\epsilon_+$,
which, at $N>N_c=(2j_0+1)/(U-W)$, yield
$N_0=\frac{1}{2}(N+N_c)$ and
$N_+=\frac{1}{2}(N-N_c)$.

\begin{figure}[t]
\includegraphics[width=0.82\linewidth,height=0.7\linewidth]{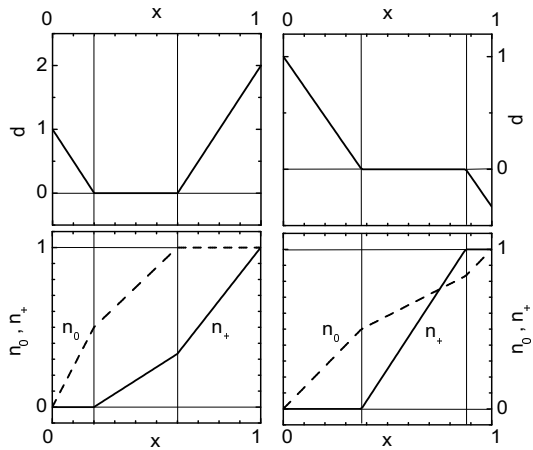}
\hskip 1.5 cm \caption{ Top panels: Dimensionless distance
$d=(\epsilon_+-\epsilon_0)/D$ between levels $+$ and $0$ as a
function of the ratio $x=N/(2j_0+2j_++2)$. Lower panels:
Occupation numbers $n_k$ for levels $0$ and $+$. Input parameters:
$U=V=3, W=1$. For the left column, the ratio $r\equiv
(2j_0+1)/(2j_++1)=2/3$; for the right, $r=3$. } \label{fig:two}
\end{figure}

Results from numerical calculations are plotted in Fig.~\ref{fig:two},
which consists of two columns, each made up of two plots.
The upper panels show the dimensionless ratio
$d(x)=\left[\epsilon_+(x)-\epsilon_0(x)\right]/D$ versus
$x=N/(2j_0+2j_++2) \in [0,1]$.  The lower panels
give the occupation numbers $n_+(x)$ and $n_0(x)$.
We observe that there are three different regimes: in two of
them $d\neq 0$ and there exist well-defined sp excitations,
and in the third, the energies of the levels 0 and + coincide at
zero. Passage through the three regimes can be regarded as a
second-order phase transition, with the occupation number $n_+$
treated as an order parameter.

The sp levels remain merged until one of them is completely filled.
If the level $0$ fills first, as in the left column
of Fig.~\ref{fig:two}, then under further increase of $N$,
quasiparticles fill the level $+$, signaling that the distance
$d(N)$ again becomes positive.  This behavior resembles the
repulsion of two levels of the {\it same symmetry} in quantum
mechanics, although here one deals with the sp levels of
{\it different symmetry}.  In a  case where level
$+$ becomes fully occupied before the level 0 does, as in the right
column, the distance $d(N)$ becomes negative, and the two
levels just cross each other at this point.

In systems without pairing correlations, for example in atoms, a
pair of particles added to any sp level with $l\neq 0$ always has
total angular momentum $J\neq 0$ (Hund's rule), in principle
destroying spherical symmetry and lifting the $m$-degeneracy of
the sp energies $\epsilon_{km}$, thereby complicating the analysis
of merging.  Here we only sketch the final results (for details,
see Ref.~\onlinecite{haochen}).  The energy functional
\begin{equation}
E=\sum\epsilon_k(0)n_{km}+
\frac{1}{2}\sum f_{km,k_1m_1}n_{km}n_{k_1m_1}\
\label{eqf}
\end{equation}
replaces the functional (\ref{energy}), and the interaction
matrix $f_{km,k_1m_1}$ replaces the matrix elements
(\ref{mel}). The variational equations
\begin{equation}
 \mu=\epsilon_k(0)+\sum f_{km,k_1m_1}n_{k_1m_1}\  ,
\label{varf}
\end{equation}
in which the sum runs over some states of the last unfilled
shell that undergo merging, are to be solved numerically.
Merging provides  a qualitative
explanation of the fact that the chemical properties of
rare-earth elements differ little, in spite of marked
variation in atomic numbers. Such an explanation
is to some extent complementary to the textbook
argument that the collapse  of the electron
$4f$-orbital is responsible for the remarkable
similarity of the chemical properties of the rare-earth elements.

It is worth noting that if the number of equations (\ref{varf})
to be solved becomes large, reasonable results can be obtained with
the replacement of summation by integration to yield \cite{ks}
\begin{equation}
\mu=\epsilon^0_k+2\!\int\! f({\bf k},{\bf k}_1)\, n({\bf k}_1)\,
d^3k_1/(2\pi)^3 \,.
\label{fermc}
\end{equation}
As will now be demonstrated, the analysis of merging of sp levels in
finite systems helps us understand what is going on in infinite matter.
Let us consider a model of a heavy-fermion metal in which the
sp spectrum, evaluated usually in local-density approximation, is exhausted
by (i) a wide band that disperses through the Fermi surface,
and (ii) a narrow band situated  below the Fermi surface at a
distance $D_n$.  To facilitate the analysis, we assume that
only the diagonal matrix element $f_{nn}$ of the interaction
function $f$ referring to the narrow band is significant,
the others being negligible. The shift $\delta\epsilon_n$
in the location of the narrow band  due to switching on the
intraband interactions is given by a formula
$\delta\epsilon_n =f_{nn}\rho_n $ analogous to Eq.~(\ref{en1}),
where $\rho_n$ is the density of the narrow  band.
Suppose now that the correction $\delta\epsilon_n$ exceeds the distance
$D_n$. In this case, the HF scenario calls for the narrow band to be
completely emptied; but then the shift $\delta\epsilon_n$ must vanish.
To eliminate this inconsistency, it must be the case that only a
fraction of the particles leave the narrow band, in just the
right proportion to equalize the chemical potentials of the
two bands.  The feedback mechanism we have described positions
the narrow band exactly at the Fermi surface, resolving a
long-standing problem with the LDA scheme.

\section{ Quantum critical point in a homogeneous Fermi liquid}

In this section, we investigate  properties of Fermi liquids
in the vicinity of the FL quantum critical point (QCP), i.e. in a region
close to the critical density $\rho_{\infty}$ where the effective mass
$M^*$ diverges. Standard FL theory, which knows nothing about the QCP,
tells us that  properties of Fermi liquids are similar to those of
an ideal Fermi gas, with differences merely involving a numerical factor,
the ratio $M^*/M$. However, at the critical density $\rho_{\infty}$
where $M^*$ diverges, the FL spectrum
$\epsilon_{\mbox{\scriptsize FL}}(p;\rho)$ vanishes identically,
signalling that the standard FL theory requires a cure.
Remarkably, the Landau approach itself contains a medicine to treat
the disease. The cure lies in the relation \cite{lan}
\begin{equation}
\frac{\partial \epsilon(p)}{\partial {\bf p}}=\frac{{\bf p}}{M}
+2\int f({\bf p},{\bf p}_1)
\frac{\partial n(\epsilon(p_1))}{\partial {\bf p}_1}d^3p_1/(2\pi)^3 \,,
\label{lansp}
\end{equation}
which connects the quasiparticle spectrum $\epsilon(p)$ and the momentum
distribution $n(\epsilon)$ through the Landau interaction function
$f({\bf p},{\bf p}_1)$.  Referring back to the preceding section in
which we considered the merging of sp levels in finite Fermi systems,
this relation has its conceptual counterpart in Eq.~(\ref{rel1}).
From Eq.~(\ref{lansp}) one finds
\begin{equation}
M/M^*(\rho,T=0)=1-F^0_1(\rho)/3 \,,
\label{meff}
\end{equation}
having introduced the dimensionless first harmonic
$F^0_1(\rho)=f_1(p_F,p_F)N_0$, where $N_0=p_FM/\pi^2$ and
$f_1(p_F,p_F)$ is the first harmonic
of  $f({\bf p}_1,{\bf p}_2)$. From Eq.~(\ref{meff})
we infer that  realization of the divergence of $M^*$ hinges on
the presence of sufficiently large velocity-dependent components
in $f$, since one must have $F^0_1(\rho_{\infty})=3$.

\begin{figure}[t]
\includegraphics[width=0.9\linewidth,height=0.68\linewidth]{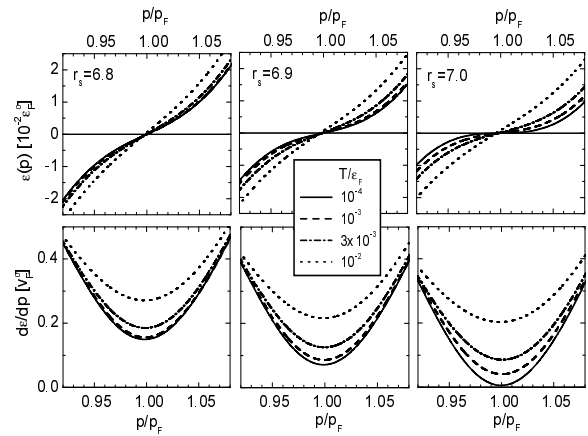}
\hskip 1.5 cm \caption{Single-particle spectrum $\epsilon(p)$ in
units of $10^{-2}\,\epsilon_F^0$ (top panels) and its derivative
$d\epsilon(p)/dp$ in units of $v_F^0=p_F/M$ (bottom panels) for
the 2D electron gas at $r_s=6.8$ (left column), $r_s=6.9$ (middle
column), and $r_s=7.0$ (right column). Spectra and derivatives are
shown as functions of $p/p_F$ at four values of temperature in
units of $\epsilon_F^0=p_F^2/2M$. } \label{fig:eg_fl}
\end{figure}

Quantitative studies of the spectrum $\epsilon(p)$ first appeared
in connection with the problem of fermion condensation (see below)
some forty years after the creation of FL theory. Close to the
critical density where $M^*(\rho)$ diverges, the group velocity
$d\epsilon(p;\rho\to \rho_{\infty})/dp$ ceases to be constant,
acquiring a parabola-like shape.\cite{zb,bz} This is seen from
Fig.~\ref{fig:eg_fl}, where the microscopically evaluated
\cite{bz} sp spectrum and group velocity of a 2D electron gas are
shown.

The structure of the sp spectrum $\epsilon(p,\rho)$ in the vicinity of
the Fermi-liquid QCP can be elucidated without resorting to the help of
a computer. Instead one expands the relevant quantities on both sides of
Eq.~(\ref{lansp}) in Taylor series, thereby obtaining
\begin{equation}
d\epsilon/dp\simeq p_F/M^*(\rho,T)+ v_2(p-p_F)^2/Mp_F  \  .
\label{gro}
\end{equation}
Inserting this expression into Eq.~(\ref{lansp}) and performing
rather lengthy algebra, one finds \cite{ckz}
\begin{equation}
 M/M^*(T,\rho_{\infty})\propto T^{2/3}\  .
\end{equation}
Thus, within the Landau approach, the density of states diverges as
$T^{-2/3}$ when the density $\rho$ approaches $\rho_{\infty}$.
The corresponding asymptotic behaviors of the spin susceptibility
and the ratio $C(T)/T$
are
$\chi(T,\rho_{\infty})\propto C(T,\rho)/ T\propto T^{-2/3}$,
while the entropy $S(T,\rho_{\infty})$ and the thermal expansion
$\beta(T,\rho_{\infty})$ behave as $T^{1/3}$.

Imposition of a static external magnetic field $H$ brings into play
a new dimensionless parameter $R=\mu_BH/T$ and opens another arena for
testing the Landau approach. The function $n(\epsilon(p))$ entering
Eq.~(\ref{lansp}) is then replaced by
$\left[n(\epsilon_+(p))+n(\epsilon_-(p)\right]/2$, where
$n\left(\epsilon_{\pm}(p)\right)=
\left[1+\exp(\epsilon(p)/T \pm R/2))\right]^{-1}$.
Proceeding as before, one uncovers a scaling behavior
\begin{equation}
M/ M^*(T,H,\rho_{\infty})\propto T^{2/3}a(R)
\label{meh}
\end{equation}
of the effective
mass,
where $a(R)$ has been evaluated in a closed form in Ref.~\onlinecite{ckz}.
In the limit $T\to 0$ or equivalently $R\to \infty$, this behavior
 simplifies, yielding  the analytic form
\cite{shag1,ckz} $M^*(T=0,H,\rho_{\infty})\propto H^{2/3}$ for the effective mass.
Thus, on the metallic side of the phase transition associated with
the divergence of the effective mass, imposition of a static magnetic
field satisfying $\mu_BH>T$ renders the effective mass
$M^*(T,H,\rho_{\infty})$ finite, promoting the recovery of Landau FL
theory. This behavior, consistent with the experiment, remains elusive
 in any approach involving  spin fluctuations as a basic
ingredient.

\begin{figure}[t]
\includegraphics[width=0.9\linewidth,height=0.9\linewidth]{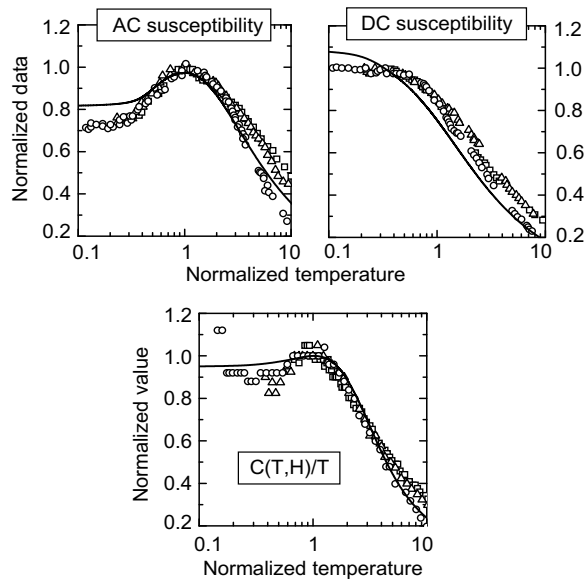}
\hskip 1.5 cm \caption{ op panels: Normalized magnetic
susceptibility $\chi(T,H)/\chi(T_P)$ (top-left panel) and
normalized
 magnetization ${\cal M}(T,H)/{\cal M}(T_P)$ (top-right panel) for
CeRu$_2$Si$_2$ in magnetic fields 0.20 mT (squares), 0.39 mT
(triangles), and 0.94 mT (circles), plotted against normalized
temperature $T/T_P$ (Ref.~\onlinecite{takahashi}), where $T_P$ is
the temperature at peak susceptibility. The solid curves trace the
universal behavior predicted by the present theory. Bottom panel:
The normalized ratio $C(T,H)T_M/C(T_M)T$ for
YbRh(Si$_{0.95}$Ge$_{0.05}$)$_2$ in magnetic fields 0.05 T
(squares), 0.1 T (triangles), and 0.2 T (circles), versus the
normalized temperature $T/T_M$ (Ref.~\onlinecite{custers}), where
$T_M$ is the temperature at maximum ratio $C(T,H)/T$. The solid
curve shows the prediction of our theory.}
\label{fig:sca}
\end{figure}

Along the same lines, one finds that close to the QCP,
the magnetic moment and AC spin susceptibility display a scaling
behavior. Following Ref.~\onlinecite{takahashi}, Fig.~\ref{fig:sca} presents
the results of numerical calculations of these quantities as
functions of the normalized temperatures $T/T_P$ and $T/T_M$.
We see that the model developed here reproduces the experimental
scaling behaviors of both the spin susceptibility \cite{takahashi}
of the heavy-fermion metal CeRu$_2$Si$_2$ and the specific heat
\cite{custers} of the heavy-fermion compound
YbRh(Si$_{0.95}$Ge$_{0.05}$)$_2$, {\it without any adjustable
parameters}.

\section{Going beyond the point of fermion condensation}

For decades, there has been virtually universal acceptance of Landau's
hypothesis that in homogeneous systems of fermions there exists
a one-to-one  $T$-independent correspondence between the
sp spectrum $\epsilon(p)$ and the momentum $p$. This postulate
(see the text accompanying formula (4) in Ref.~\onlinecite{lan}),
implying that the quasiparticle group velocity $d\epsilon/dp$
at the Fermi surface is practically {\it independent of $T$}, has
been regarded as a cornerstone of FL theory. However, numerical
investigations of Eq.~(\ref{lansp}) performed in recent years
\cite{zb} have demonstrated that this hypothesis is incorrect in
the vicinity of the QCP. This flaw was in fact uncovered already
in 1990's in connection with an analysis of the necessary condition
for stability of the ground state with respect to the rearrangement
of the ground-state momentum distribution $n(p)$. The pertinent
stability condition reads
\begin{equation}
\delta E_0(n)=2\int \epsilon(p)\delta n(p)d^3p/(2\pi)^3\geq 0 \ ,
\label{infl}
\end{equation}
thus requiring nonnegativity of the variation of the ground-state energy
$E_0$ under any admissible variation of $n(p)$. In the case of the Landau
state where $n(p)$ is simply the Fermi step $n_F(p)$, this implies that
the sign of $\epsilon(p)$ must coincide with the sign of the difference
$p-p_F$. However, it turns out that there exists a class of NFL
solutions having a completely flat portion $\epsilon(p)=0$ in a region
embracing the Fermi surface,\cite{ks,vol,noz,physrep} which in principle
provides a different minimum of the functional $E_0(n)$. Flattening of
the sp spectrum  in this way is inescapably accompanied by another
peculiarity: occupation numbers $n_*(p)$ in the flat region differ
from 0 and 1.
As a result, the Fermi surface swells from a surface to a volume in
3D and from a line to a surface in 2D. The phenomenon of swelling
of the Fermi surface, first documented in Ref.~\onlinecite{ks}
and called fermion condensation, is its distinctive signature.
We have already encountered swelling of the Fermi surface in Sec.~2,
where the merging of sp levels was studied.  The set of states with
$\epsilon({\bf p})=0$ is called the {\it fermion condensate} (FC)
in transparent analogy with the boson condensate that forms at $T=0$
in a Bose liquid, and in which macroscopically many particles have
energy coincident with the chemical potential $\mu$.

The existence of ground states with occupation numbers $0<n_*(p)<1$
follows directly from the fundamental idea of the FL approach that $E_0$
is a functional of $n(p)$. To see this more clearly, it is instructive
to invoke a mathematical correspondence with the functional $E_0(\rho)$
of statistical physics. If the interaction is small,
the latter attains its minimum for solutions describing {\it gases}.
However, if the interaction between the particles
is sufficiently strong, then nontrivial solutions of the variational
condition $\delta \Omega/\delta \rho=0$ on the thermodynamic
potential $\Omega=E_0-\mu N$ describe {\it liquids}.
From a mathematical point of view, the functional $\Omega(n)$
must have two types of solutions.  Assuredly, in weakly interacting
systems, its minimum is attained at $n_F(p)=\theta(p-p_F)$, which
minimizes the kinetic energy alone. On the other hand, if the
potential energy becomes strong, there arise nontrivial solutions
of the variational condition
\begin{equation}
\frac{\delta \Omega}{\delta n({\bf p})}=0 \ , \quad {\bf p}\in C \,,
\label{varh}
\end{equation}
over a {\it finite domain $C$ adjacent to the Fermi surface}.
\cite{ks,noz,physrep,vol}  In Landau's formulation of Fermi-liquid
theory, at $T=0$ the derivative $\delta E/\delta n({\bf p})$ is just
the sp energy, while $\delta \Omega/\delta n({\bf p})$ is the sp
energy $\epsilon({\bf p})$ measured from the chemical potential $\mu$.
Thus we arrive at the relation
\begin{equation}
\epsilon({\bf p})=0 \  , \quad {\bf p}\in C \,,
\end{equation}
characterizing the fermion condensate.

In contrast to other second-order phase transitions, no symmetry
is violated in the FC phase transition, and therefore the choice
of the associated order parameter is not immediate. The FC density
$n_*(p)$ can be treated as such an order parameter. Upon rewriting
Eq.~(\ref{dist}) as
$\epsilon({\bf p},T)=T\ln [(1-n({\bf p}))/n({\bf p})]$
and inserting into this expression of the FC solution $n_*({\bf p})$,
one finds
\cite{noz}
\begin{equation}
\epsilon({\bf p},T\to 0)=T\ln \frac{1-n_*({\bf p})}{n_*({\bf p})} \ ,
\quad {\bf p}\in C \  .
\label{spet}
\end{equation}
Thus at low $T$, the energy degeneracy of FC  is lifted,
the FC plateau in $\epsilon({\bf p})$ being inclined with
a slope proportional to temperature. \cite{noz}

This salient feature of the FC spectrum exhibits itself in the
magnetic susceptibility $\chi(T)$, providing Curie-Weiss behavior
$\chi(T)\propto 1/T$ in normal states of systems with a FC.
Indeed, upon inserting the function $n_*({\bf p})$ into the standard
FL relation
$$
  \chi_0=-2\mu_e^2\!\int\!\frac{dn(\epsilon_{\bf p})}{d\epsilon_{\bf p}}\,
  d^3p/(2\pi)^3 $$
\vspace{-.5truecm}
\begin{equation}
\equiv \frac{2\mu_e^2}{T}\!
  \int\! n({\bf p})\,[1-n({\bf p})]\,d^3p/(2\pi)^3 \ ,
\label{chi0}
\end{equation}
we find a Curie contribution to the spin susceptibility\cite{zk}
given by
\begin{equation}
  \chi_*(T)=\frac{\kappa\,\mu_e^2}{T} \ , \quad
  \kappa=2\!\int\! n_*({\bf p})\,[1-n_*({\bf p})]\,d^3p/(2\pi)^3 \ .
\label{Curie}
\end{equation}
The effective Curie constant
in Eq.~(\ref{Curie}) {\it is reduced by the dimensionless parameter
$\kappa$ relative to the standard Curie law} $\chi_0=\mu_e^2/T$.
Accounting for the spin interaction amplitude $g_0$ generates the
Curie-Weiss law $\chi(T)=\mu_e^2\kappa/(T-\Theta_W)$ with a Weiss
temperature $\Theta_W=g_0\kappa\mu_e^2$. Such a behavior of $\chi(T)$
is observed in $^3$He films, where the low-$T$ Curie constant is
about 4 times smaller than that for high $T$, as shown in Fig.\ 1 of
Ref.~\onlinecite{godfrin}, which gives $\kappa\approx0.25$ in this case.

Another important feature inherent in systems having a FC stems from
the fact that the variational condition (\ref{varh}) holds at finite
temperatures,\cite{noz} provided the thermodynamic potential is redefined
according to the conventional relation $\Omega=E-\mu N-TS$, where
the entropy $S$ is given by the Landau formula
$$
S=-2\int [ n({\bf p})\ln n({\bf p}) +
        (1-n({\bf p}))\ln (1-n({\bf p})) ]
        $$
\vspace{-.5truecm}
\begin{equation} \times   d^3p/(2\pi)^3 \ .
\label{entr}
\end{equation}
Upon inserting here the distribution $n_*(p)$, we infer that systems with
a FC possess {\it a finite, independent of} $T$ {\it entropy} $S_*$.
\cite{ks,physrep,noz} The entropy-excess value, proportional to the total
FC density, changes gradually under variation of input parameters.
Although it does not contribute to the specific heat, it produces an
enormous enhancement of the thermal expansion coefficient
$\beta=\partial V/\partial T\equiv -\partial S/\partial P$ and
the Gr\"uneisen ratio $\Gamma=\beta/C$ \cite{zksb}.
Experiment \cite{oeschler} shows
that in normal states of  several heavy-fermion metals, $\beta$ is
in fact temperature-independent at low $T$ and exceeds
typical values for ordinary metals by a factor $10^3$--$10^4$.
With $\beta\to\rm const$ and $C(T)\to0$, the Gr\"uneisen ratio
$\Gamma=\beta/C$ diverges at low $T$, as is observed
experimentally \cite{gruneisen}.

It should be emphasized that the existence of the residual entropy
$S_*$ at zero temperature contradicts the third law of thermodynamics
(the Nernst theorem). To ensure that $S=0$ at $T=0$, localized spins
are known to order magnetically due to spin-spin interactions.
Similarly, a system with a FC must experience some sort of
low-temperature phase transition eliminating the excess entropy $S_*$,
e.g.\ the second-order phase transition to a superconducting state
\cite{ks}. The presence of the FC in the ground state exhibits itself
in a jump $\Delta C\approx 4.7\,\kappa$ of the specific heat at $T_c$,
governed by the FC parameter $\kappa$ in the
Curie law\cite{prl} (\ref{Curie}).
Thus, the ratio $\Delta C/C_n$ can be very high when $T_c$ is low,
because $C_n\to0$ as $T\to0$ while $\Delta C$ remains finite. Such
a situation is encountered for example in the heavy-fermion metal
CeCoIn$_5$, with $T_c=2.3 K$ and $\Delta C/C_n\approx 4.5$, over three
times higher than the BCS value.\cite{petrovic}

\section{Discussion}

In any conventional homogeneous Fermi liquid, e.g.\ liquid $^3$He,
the momentum ${\bf p}$ of an added particle can be associated with
a certain quasiparticle. Similarly, in most spherical odd nuclei,
the total angular momentum $J$ in the ground state is carried by
an odd quasiparticle. In atomic physics, the electronic configuration
of ions of elements belonging to the principal groups of the periodic
table repeats that of preceding atoms. By contrast, beyond the point
of fermion condensation in homogeneous matter or the point of merging
of sp levels in finite systems, the ground state features
a multitude of quasiparticle terms and therefore exhibits
a different, more complicated  character, as in the comparison of
a chorus with a dominant soloist.

Notwithstanding evident commonalities, there is a considerable
difference between conditions for the ``level-mergence''
phenomenon in homogeneous Fermi liquids and in finite Fermi
systems with degenerate sp levels. In the former, the presence
of a significant velocity-dependent component in the interaction
function $f$ is needed to promote fermion condensation, while
in the latter, sp levels can merge even if $f$ is momentum-independent.
The reason for this difference is simple: in the homogeneous case,
the matrix elements $u$ and $w$ of Eq.~(\ref{mel}) are equal to
each other, implying zero energy gain due to the rearrangement when
velocity-dependent forces are absent.

The salient feature of the fermion condensation phenomenon
discussed in this paper is the flattening of the quasiparticle
dispersion $\epsilon({\bf p})$ at
the Fermi level on the ordered side of the driving phase
transition. Prime consequencies of this flattening are:
(i) the magnetic susceptibility follows the Curie-Weiss law with an
effective Curie constant proportional to  the FC parameter $\kappa$,
(ii) the entropy has a temperature-independent term $S_*$ that
greatly increases the thermal expansion coefficient
$\beta=-\partial S/\partial P$ at low $T$,
(iii) the excess entropy $S_*$ released below the superconducting
transition temperature $T_c$ dramatically reduces $\beta$, enhancing
the specific-heat jump $\Delta C/C_n$. All these features go
unexplained in any other microscopic approach.

On the disordered side of the transition, close to the QCP,
the proposed scenario adequately explains the low-$T$ data on
the spin susceptibility, predicting $\chi^{-1}(T)\sim T^{\alpha}$
in the critical density region with a critical exponent
$\alpha\simeq 2/3$.  The spin-fluctuation model (SFM) fails
to produce $\alpha<1$. Further, our scenario explains the scaling
behavior $\chi^{-1} \sim T^\alpha F(H/T)$ of the spin susceptibility
in static magnetic fields, whereas the SFM fails to do so.
Finally, within the scenario advanced here, FL behavior is recovered
close to the QCP by imposing a tiny magnetic field satisfying
$\mu_BH>T$. In the SFM there is no such provision for reinstating
FL theory.

\begin{figure}[t]
\includegraphics[width=0.8\linewidth,height=0.6\linewidth]{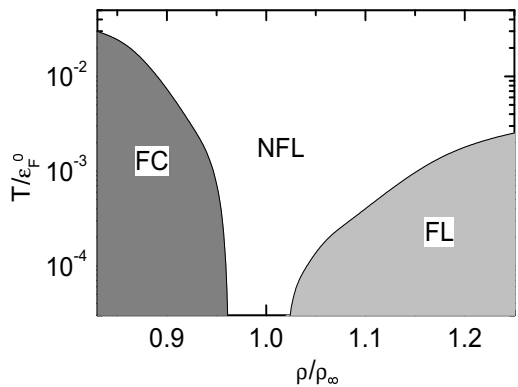}
\hskip 1.5 cm \caption{ Phase diagram of 2D electron gas in
$(T,\rho/\rho_{\infty})$ variables. Curves show crossovers between
usual Fermi liquid (FL), fermion condensate (FC) and non-Fermi
liquid (NFL) phase with the critical index $\alpha\simeq 0.6$. }
\label{fig:ph_dia}
\end{figure}

\section{Conclusion}
The publication fifteen years ago of the first article \cite{ks} on
the theory of fermion condensation, in which NFL behavior was deduced
{\it from the Landau approach itself}, triggered a wave of the criticism
and disbelief.  The judgment, ``This theory is an artifact of the Hartee-Fock
method''  was typical.  By now the debates on this subject have become
pointless: numerical solution of the Landau equation (\ref{lansp})
is the best way to calm down the critics.  Due to lack of space, we
present here only a phase diagram of the two-dimensional homogeneous
electron gas calculated from Eq.~(\ref{lansp}). This phase diagram,
as plotted in Fig.\ \ref{fig:ph_dia},
essentially coincides with predictions of the theory of fermion
condensation.  Distinctions occur only at extremely low temperatures
in the region adjacent to the QCP (for details see Ref.~\onlinecite{zb}).
Thus, it is fair to say that the theory of fermion condensation is just
another chapter of Fermi-liquid theory.  Accordingly, we abandon the
conventional attitude of standard FL theory that properties of Fermi
liquids are always similar to those of an ideal Fermi gas, since as
we have seen, there exists a whole region of densities where this is
not the case.  Still to be answered is the important question: is the
Fermi-liquid approach \cite{lan} relevant to the experimental situation
in strongly correlated Fermi systems close to the QCP and beyond it?
With the passage of time, a positive answer seems to be more and more
evident.

We thank A.~Chubukov, P.~Coleman, P.~Fulde, H.~Li, K.~Kikoin,
G.~Lonzarich, J.~Thompson, J.~Saunders, F.~Steglich, and G.~Stewart
for fruitful discussions.  This research was supported by
Grant No.~NS-8756.2006.2 (VAK and MVZ) from the Russian Ministry
of Education and Science and the McDonnell Center for the
Space Sciences (VAK).

\end{document}